\documentclass[12pt]{article}
\usepackage{amsmath}
\usepackage{a4}
\usepackage{epsfig}
\usepackage{dcolumn}
\usepackage{rotating}

\begin{document}

\newcommand{\av}[1]{\left\langle#1\right\rangle}
\newcommand{\ns}{N_S}
\newcommand{\nt}{N_\tau}
\newcommand{\ds}{\Delta S}
\newcommand{\po}{P_0}
\newcommand{\ps}{P_S}
\newcommand{\sqs}{\sqrt {\sigma}}
\newcommand{\pt}{P_\tau}
\newcommand{\sz}{{\sigma_0}}
\newcommand{\tr}{\operatorname{Tr}}
\newcommand{\re}{\operatorname{Re}}

\setlength{\topmargin}{-12mm}

\def\be{\begin{equation}} 
\def\ee{\end{equation}} 
\def\bea{\begin{eqnarray}}
\def\eea{\end{eqnarray}}

\renewcommand{\thefootnote}{\alph{footnote}}
\begin{flushright}
\parbox{3cm}{BI-TP 2009/18\\
HU-EP-09/45}
\end{flushright}
\vspace{5mm}

\begin{center}

{\Large\bf Three dimensional finite temperature SU(3) gauge theory   
in the confined region and the string picture}
\\[1.8ex]
{\bf P. Bialas$^{1,2}$\footnote{pbialas@th.if.uj.edu.pl},
  L. Daniel$^1$\footnote{daniel@th.if.uj.edu.pl},
    A. Morel$^3$\footnote{andre.morel@cea.fr}, 
    B. Petersson$^{4,5}$\footnote{bengt@physik.uni-bielefeld.de}}
    \\[1.2mm]
    $^1$ Inst. of Physics, Jagellonian University\\
    ul. Reymonta 4, 30-059 Krakow, Poland \\[1.2mm]
    $^2$ Mark Kac Complex Systems Research Centre\\
    Jagellonian University, Reymonta 4, 30--059 Krakow, Poland\\[1.2mm]
    $^3$Institut de Physique Th\'eorique de Saclay, CE-Saclay \\
    F-91191 Gif-sur-Yvette Cedex, France\\[1.2mm] 
    $^4$ Fakult\"at f\"ur Physik, Universit\"at Bielefeld \\
    P.O.Box 10 01 31, D-33501 Bielefeld, Germany \\[1.5ex]
    $^5$ Humboldt-Universit\"at zu Berlin, Institut f\"ur Physik, \\
    Newtonstr. 15, D-12489 Berlin, Germany
    \end{center}

\vspace{1.5cm}

\begin{abstract}

\medskip
\noindent
We determine the correlation between Polyakov loops in three
dimensional
SU(3) gauge theory in the confined region at finite temperature.
For this purpose we perform lattice calculations
for the number of steps in the temperature direction equal to six. This
is
expected to be in the scaling region of the lattice theory. We compare
the
results to the bosonic string model. The agreement is very good for
temperatures
$T<0.7\;T_c$, where $T_c$ is the critical temperature. In the region
$0.7\;T_c<T<T_c$ we enter the critical region, where the critical
properties of the correlations are fixed by universality to be
those of the two dimensional three state Potts model.
Nevertheless, by calculating the critical lattice coupling,
we show that the ratio of the critical temperature to the square root of
the
zero temperature string tension, where the latter is taken from the literature, remains very
near to the string model prediction.
\end{abstract}

\section{Introduction}

\medskip
\noindent
Three dimensional SU(3) gauge theory has many properties in common with the
corresponding four dimensional theory. At zero temperature lattice calculations
show the
confinement of heavy quarks in a linear potential. Furthermore, they
predict a mass gap and a nontrivial glueball spectrum.  At finite 
temperature there is a deconfining phase transition. In contrast to SU(3)
in four dimensions the transition is second order. 
Lattice calculations of the critical
indices are consistent with the transition being in the universality class
of the two dimensional three state Potts model \cite{lego}, as 
expected from general arguments\cite{svetitsky}.

\medskip
\noindent
In a previous paper, we have studied in detail
the equation of state in the high temperature phase by performing
extensive lattice calculations, as is possible in three
dimensions, and thus obtaining very precise results
\cite{bialas}.
In this note we describe a study of the theory in the confined
phase below the transition.  We work at $\nt=6$, which from the experience
gathered in \cite{bialas} we believe is in the scaling region. Thus
our result should be relevant for the continuum limit. In particular we calculate
the correlations between Polyakov loops.  It has been proposed long
ago that these can be described by an effective string model, the
Nambu-Goto bosonic string
\cite{pisarski,olesen}. In this model the temperature dependent
 string tension becomes zero at a
critical temperature $T_c$ corresponding to the Hagedorn temperature
of the string theory. The critical index for the approach of the string
tension to
zero is $\nu=1/2$. This index is different from that of the two dimensional
three state Potts model, which has $\nu=5/6$. In fact the above
predictions from the string model are independent
of the gauge group, but only depend on the dimensionality of space.
Early investigations have shown that nevertheless the critical temperature
is not far from the prediction of the Nambu-Goto model
\cite{lego,teper,liddle}. It has also been shown that for low enough
temperature the correlations between Polyakov loops are very well described
by the effective string model \cite{anthen}.
Finally, field theory investigations have proven that the first three 
terms of an expansion in $T^2$  
of the string model are in fact universal results for a fluctuating
bosonic string \cite{luscher,aharony}. But the expansion of course does not tell 
anything about the existence of a singularity at $T_c$. 

\medskip
\noindent
Here, from the lattice, we extend the earlier results on the determination 
of the ratio between the critical temperature and the square root 
of the string tension.
Furthermore we study in detail the behaviour of the Polyakov loop
correlations
in the region $T_c/2<T<T_c$. For this purpose, we have taken data for a 
dense set of temperatures, varying the temperature by changing the lattice
coupling constant.
This method is complementary to the one used in
 \cite{anthen}, 
where the lattice extent in the
temperature direction is varied. As will be shown,
where our results overlap with those of \cite{anthen}, 
they agree with each other.
It is also very interesting to study in detail the critical region,
near $T_c$, where we do not expect the string model to describe
the correlations. Instead, in fact, critical scaling is found to be 
a very efficient tool to describe the correlation length between Polyakov
loops.
We postpone the discussion of
these issues to a separate publication \cite{bialas2}.

\medskip
\noindent
In the next section we present the lattice setup and define the
quantities of interest in the string context. The simulations are
described in section 3, together with the determination of the critical
temperature for a lattice extent in the temperature direction 
equal to 6.
Section 4 is devoted
to the comparison of the results obtained with the string picture, by
reference to the Nambu-Goto model and to field theoretic
approaches. 
 A summary and conclusions are
given in a last section.

\medskip
\noindent

\section{Polyakov loop correlations at finite temperature. Lattice
setup.}

\medskip
\noindent
We simulate the three dimensional $SU(3)$ theory, regularized
on a finite euclidean lattice with lattice spacing $a$, $\nt$ points in the (inverse)
temperature direction, defined as the 0 direction,
$\ns$ points in the two space directions $1,2$ and periodic 
boundary conditions in all directions. We use the standard Wilson action:
\be\label{wilson}
S(U_P) = \beta \sum_P (1-\frac{1}{3}\re TrU_P),
\ee
\medskip
\noindent
where P denotes one of the $3\nt \times \ns^2$ plaquettes on the
lattice, $U_P$ is the product of the $U$-matrices around the
plaquette, and $\beta$ is the lattice coupling constant. From the
classical limit of the lattice action we may write

\be
\beta= \frac{6}{ag^2}
\ee

\medskip
\noindent 
where $g^2$ is the (dimensionful) gauge coupling constant of the continuum
theory.
We define the temperature and the volume of the lattice by

\bea
\frac{1}{T} & = & a\nt, \label{temp}\\
V & = & (a\ns)^2. \label{vol}
\eea

\medskip
\noindent
In the confined phase, we measure the correlations between Polyakov 
loops winding around the temperature direction:
\begin{equation}\label{eq:polydef}
 L(x_1,x_2)=\tr\prod_{n=0}^{N_\tau-1} U_{\tau} \left(\vec{x}+ n\,\hat{e}_{\tau} \right),
\end{equation}
where $U_\tau$ denotes the link in the time direction, whose origin is
located in space at 
$\vec{x}=(x_1,x_2)$ and $\vec{e}_\tau$ is the unit vector in the
time direction. In 2-dimensional space, we define the correlation function
$G(z)$ between two lines at distance $z$ from each other by

\begin{equation}\label{eq:polya}
\begin{split}
G(z)\equiv\frac{1}{2 N^3_S}\Biggl(&\sum_{x_1,x_2,x'_2}
\re\av{L(x_1,x_2)L^\ast(x_1+z,x'_2)}+ \\
&\sum_{x_1,x'_1,x_2}
\re\av{L(x_1,x_2)L^\ast(x'_1,x_2+z)} \Biggr)
\end{split}
\end{equation}

Throughout the paper, the symbol $\av{...}$ denotes an average over
a set of gauge configurations.
Due to periodic boundary conditions, any coordinate is understood 
modulo $N_s$, and all sums run over intervals of length $\ns$.
Using the discrete lattice symmetries, we may rewrite $G(z)$ as 
\begin{equation}
G(z)=\frac{1}{N_s}\sum_{x_2}\re\av{L(0,0)L^\ast(z,x_2)}
\end{equation} 
In the forthcoming analysis, it will be assumed that for $z$ large
enough (see details in section 3), contributions from excited states
of the 2-dimensional system of Polyakov loops can be neglected, and 
$G$ represented by the contribution of its ground state
energy only. In this situation, we will use the parametrization

\begin{equation} \label{Polya}
G(z)=
b\,\cosh\left(m (\frac{\ns}{2}-z)\right)\,+\,c. 
\end{equation}
\medskip
\noindent
Using (\ref{temp}), the ground state energy, here denoted $m$ in lattice units, 
is related to its physical value $M$ by
\be
m=Ma=\frac{1}{\nt}\,\frac{M}{T}. \label{ma}
\ee
In the right hand side of (\ref{Polya}), $b$ and $c$ are assumed to be
constants for given lattice parameters. 
The first term is the 
lattice expression of $G$ for a free bosonic field, whose normalization
determines $b$. In our region of investigation the constant c is essentially 
consistent with zero, and always negligible in practice.

From now on, we will interpret $M(T)$  
as the ground state energy of a flux tube, writing 

\be
\frac{M(T)}{T}=\frac{\sigma(T)}{T^2}  \label {stringt}
\ee

\medskip
\noindent
where $\sigma(T)$ is a temperature dependent string tension. We keep
this definition of $\sigma(T)$ for any $T$, and from
Eqs. (\ref{temp},\ref{ma}) rewrite the latter equation as 
\be
\frac{\sigma(T)}{T^2}\,=\,m\nt. \label{sigm} 
\ee 
In order to study the
temperature dependence of $\sigma(T)$, and compare our data with the
string picture it is convenient to use the zero temperature string
tension $\sz$ as a scale, i.e. using Eqs. (\ref{temp},\ref{sigm}) to
write 
\bea
\frac{T}{\sqrt{\sz}}\,=\,\frac{1}{\nt\,a\sqrt{\sz}}  \label{tsig}\\
\frac{\sigma(T)}{\sz}\,=\,\frac{m}{\nt\,(a\,\sqrt{\sz})^2} \label{sigs0}
\eea 
The quantity $a\sqrt{\sz}$ has been recently measured from
numerical simulations using large $\ns^3$ lattices \cite
{liddle,anthen}, leading to high accuracy results at $\beta=$ 8.1489,
11.3711, 14.7172, 18.131, 21 and 40. Because we need it for a dense
set of $\beta$ values in the 10 to 22 range, we determined
$a\sqrt{\sz}$ for any $\beta$ by fits to the above numerical data.
After some trials, we finally retained the following parametrization
\bea
F_{\sz}(\beta)\,&\equiv&\,a\sqrt{\sz}\,=\,\frac{h}{\beta}\,\frac{\beta-z}{\beta -p}. \label{fsz}\\
h=3.3257\quad\quad z&=&1.99 \quad \quad p=3.69 
\eea 
Not only all the
data points are perfectly fitted, but it is interesting to notice that
the zero and pole positions $z$ and $p$ in $F_{\sz}$ suggest the
existence of a cross-over from a weak to a strong coupling regime, for
$\beta$ of the order of a few units, well below the lowest $\beta$
value needed. Moreover, we use this function only for interpolating
accurate data. We do not quote the errors on the fitted parameters, as
they are anyway highly correlated. What is important is the error in the
interpolating function.   We found that the absolute error on
the inverse of the function $F_{\sz}(\beta)$ is approximately constant
and equal to $0.002$. 

\section{Simulations and results}

The simulations were done in a standard way using the Wilson action
\eqref{wilson}.  We used one heathbath sweep followed by four
overrelaxation sweeps. The Polyakov loops correlations were measured
according to formula \eqref{eq:polya}. The measurements were performed
every five combined (heathbath and overrelaxation) sweeps.  At least
20,000 measurements were kept for every $\beta$ and $N_S$ combination,
their average number being of the order of 60,000. These measurements are not
independent, and  the autocorrelation time for the correlation functions was
found to be 10 in the worst case ($\beta=20$), resulting in a minimum of 2,000
independent measurements
($\beta=20, N_s=96$).  The ground state energies $m$ were obtained by fitting
formula \eqref{Polya}, and their errors estimated 
using the blocked boostrap method as now described.

The total set of measurements for given values of $\beta$ and lattice size
was first divided into blocks of 500 measurements, in order to take care
of the autocorrelation time. Then 100 bootstrap samples were generated, each one
resulting from drawing a new set of blocks randomly from the original
sample. For each bootstrap sample we calculated the average value
of the correlation function. The average of those averages was taken
as the final value of the correlation function,  and their standard
deviation as the corresponding error.

The errors on the fitted parameters were obtained by repeating this
procedure once more. For each bootstrap sample, we 
fitted  formula \eqref{Polya} for $z$ in an interval $[z_0,N_s/2]$, using the
$\chi^2$ value calculated from the errors found in the
previous step. We so obtained 100 sets of fitted parameters $m,b,c$
and used the standard deviation between them as an estimate of their errors.

Eq.\eqref{Polya} retains the contribution from the
ground state only. In order to control the possible effect of  higher
states, we measured the dependence of the fitted masses on the
$z_0$ parameter, and found that 
the systematic error associated with the choice of $z_0$ becomes  smaller
than the statistical error for $z_0 \ge 4$.  In 
table~\ref{tb:amc6}, we list the final values and statistical errors of 
the parameters.

The value $\beta_c$ of the lattice coupling at which the transition takes place
has been estimated using the Binder cumulant approach \cite {binder}, following 
the method described in \cite{binder-cumulants} and applied to
the variable $|L|^2$, 
where
\begin{equation}
L\,=\,\frac{1}{\ns^2}\,\sum_{x_1,x_2}L(x_1,x_2).
\end{equation}

Using this method,  we obtain

\begin{equation}
\beta_c=21.36\pm 0.01^{(st.)}\pm 0.05^{(sys.)}
\end{equation}

The systematic error quoted covers small differences observed 
if other methods are investigated and uncertainties due to unknown subleading terms
in the scaling ansatz.
These issues will be discussed in detail in the forthcoming paper on 
the critical region \cite{bialas2}. 
Here, the present systematic error does not affect our result on 
$T_c/\sqrt{\sigma(0)}$, 
which is the relevant quantity in the forthcoming comparison with the string model.
Indeed, from the result quoted above and Eq. \ref{tsig}, we obtain 

\begin{equation}
\frac{T_c}{\sqrt{\sigma(0)}}=0.976(15), \label{toss}
\end{equation}

\noindent
where most of the error comes from the error on the function $F_\sigma(\beta)$.

\section {The string picture} 

In a fundamental paper, Pisarski and Alvarez\cite{pisarski} long ago related the
finite temperature behaviour of gauge theories in the confined region
up to the phase transition to the finite temperature behaviour of a bosonic string  
randomly fluctuating in the transverse directions. If the string is supposed to be
the Nambu-Goto string, the formula for the temperature dependent string tension
is
 
\be 
\frac{\sigma(T)}{\sz}\,=\,\sqrt{1-\frac{\pi (D-2)
    T^2}{3\sz}}, \label{Nambu} 
\ee 

\noindent
where D is the space-time dimension of the gauge theory. As we can see
from the formula, the temperature dependent string tension is not dependent
on the gauge group, but only on the number of transverse dimensions. The
string tension goes to zero at a critical temperature

\be
T_c=\sqrt{3\sigma_0/\pi} \label{tcng} 
\ee

\noindent
which coincides with the Hagedorn temperature of the string model.
The approach to this singularity has the mean field behaviour with the
exponent $\nu=1/2$.

Later, it was shown that in any dimension, the first term in the expansion in
$T^2/\sqrt{\sigma_0}$ is universal in any dimension\cite{luscher}, i.e.
 
\be 
\frac{\sigma(T)}{\sz}\,=\,1-\frac{\pi (D-2)
    T^2}{6\sz}+... \label{lusch} 
\ee 

\noindent
In three dimensions there is a special situation, as here the first three
terms are universal and coincide with the developement of the expression
for the Nambu-Goto string truncated to this order\cite{lusch2, aharony}: 
 
\be 
\frac{\sigma(T)}{\sz}\,=\,1-\frac{\pi 
    T^2}{6\sz}-  
 \frac{\pi^2 T^4}{72\sz^2}- 
 \frac{\pi^3 T^6}{432\sz^3}+...  \label{Aha}
\ee 

\noindent
Of course, this polynomial expression contains no information about a singularity
where $\sigma(T)$ vanishes, such as that of the full Nambu-Goto expression.

\begin{figure}
\begin{center}  
\includegraphics[width=14cm]{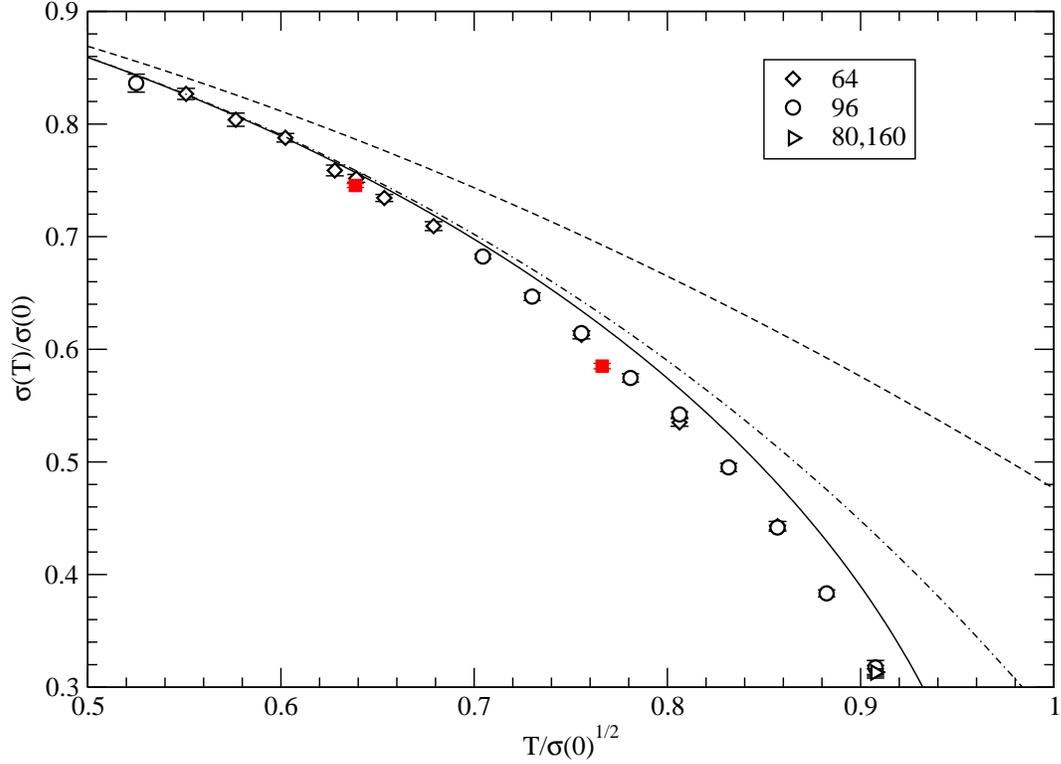}
\end{center}
\caption{\label{fig:string-tension} Finite temperature string tension
  in units of the zero temperature string tension as a function of 
  temperature for various $\ns$ values. Solid red squares
  come from \cite{anthen}. The solid line represents the Nambu-Goto expression 
  \eqref{Nambu}, and the dashed (dashed-dotted) line its expansion to the first 
  (third) order in $T^2$.  }
\end{figure}

In figure~\ref{fig:string-tension} we show the quantity $\sigma(T)/\sz$ versus
$T/\sqrt{\sz}$ as measured on the lattice using
Eqs. (\ref{tsig},\ref{sigs0},\ref{fsz}). The data points correspond to
various $\ns$ values and cover the domain $0.5< T/\sqrt{\sz}<1$.
We do not extend the domain to smaller T, because we may enter the strong
coupling region. In the figure we have also included the two points from \cite{anthen},
which are in this region. As one can see our data agree with these points. 
\begin{table}
\begin{center}
\begin{tabular}{||D{.}{.}{4}||D{.}{.}{4}||r||D{.}{.}{8}|D{.}{.}{6}l|D{.}{.}{10}||}
\hline\hline
\multicolumn{1}{||c||}{$\beta$} &\multicolumn{1}{c||}{$\frac{T}{\sqrt{\sigma(0)}}$} &\multicolumn{1}{c||}{$N_S$} & \multicolumn{1}{c|}{$m$} &\multicolumn{2}{c|}{$b$} &  \multicolumn{1}{c||}{$c$} \\
\hline\hline
12.5    & 0.5251 & 96    & 0.5055(50) & 0.10(2)         &\text{e-11}    & -0.0000045(34)         \\
                                                               
13      & 0.5509 & 64    & 0.4540(27)   & 0.26(2)       &\text{e-7}     &  0.0000031(58)         \\
13.5    & 0.5766 & 64    & 0.4029(30)   & 0.149(13)     &\text{e-6}     &  0.000005(11)          \\
14      & 0.6023 & 64    & 0.3620(18)   & 0.61(3)       &\text{e-6}     &  0.00001(1)            \\
14.5    & 0.6279 & 64    & 0.3207(20)   & 0.248(14)     &\text{e-5}     &  0.000008(20)          \\
                                                               
14.7172 & 0.639  & 64    & 0.3068(14)   & 0.404(16) 	&\text{e-5}     &  0.000008(14)          \\
15      & 0.6535 & 64    & 0.2866(12)   & 0.808(28) 	&\text{e-5}     &  0.000000(16)          \\
15.5    & 0.679  & 64    & 0.2564(14)   & 0.234(10) 	&\text{e-4}     & -0.000008(27)          \\
16      & 0.7045 & 96    & 0.22911(72)  & 0.1054(34)	&\text{e-5}     & -0.000019(11)          \\
16.5    & 0.73   & 96    & 0.2023(11)   & 0.42(2)   	&\text{e-5}     &  0.000020(23)          \\
                                                    	           
17      & 0.7555 & 64    & 0.179(1)     & 0.374(12) 	&\text{e-3}     & -0.000022(82)          \\
                                                    	           
17      & 0.7555 & 96    & 0.17942(59)  & 0.1393(37)	&\text{e-4}     & -0.000018(25)          \\
17.5    & 0.7809 & 96    & 0.157(1)     & 0.455(21)     &\text{e-4}     &  0.0000057(46)         \\
                                                    	           
18      & 0.8063 & 64    & 0.13718(87)  & 0.1783(49)	&\text{e-2}     & -0.00005(17)           \\
18      & 0.8063 & 96    & 0.13898(63)  & 0.1218(35)	&\text{e-3}     & -0.000006(48)          \\
18.5    & 0.8317 & 96    & 0.11931(87)  & 0.356(14) 	&\text{e-3}     & -0.000027(92)          \\
19      & 0.857  & 64    & 0.1005(10)   & 0.761(26) 	&\text{e-2}     &  0.00073(52)           \\
19      & 0.857  & 96    & 0.10025(46)  & 0.1023(23)	&\text{e-2}     & -0.00005(12)           \\
19.5    & 0.8824 & 96    & 0.08206(66)  & 0.2895(94)	&\text{e-2}     &  0.00010(33)           \\
                                                    	           
20      & 0.9077 & 84    & 0.0635(11)   & 0.1476(86)	&\text{e-1}     & 0.0017(18)             \\
                                                    	            
20      & 0.9077 & 96    & 0.0642(13)   & 0.846(60) 	&\text{e-2}     &  0.0033(15)            \\
                                                               
20      & 0.9077 & 160   & 0.06332(7)   & 0.699(37)     &\text{e-3}     & 0.00015(24)            \\
\hline\hline
\end{tabular}

\caption{\label{tb:amc6}The values of the  parameters in equation \eqref{Polya}
fitted to the data points shown in the figure~\ref{fig:string-tension}. }
\end{center}
\end{table}
From the figure one can also see that the NG-model is consistent with the lattice
QCD results up to about $T/\sqrt{\sz}$=0.7,  above which the corresponding curve is
definitely above the data. We also show the two curves corresponding to
Eqs. (\ref{lusch},\ref{Aha}), both higher in the graph in the whole region (since
any term of the full expansion of Eq. (\ref{Nambu}) are negative). 
The second curve cannot be distinguished
from the full NG model up to $T/\sqrt{\sz}=0.7$.
To fix the ideas, we just remark that if the last term in Eq. (\ref{Aha}) were
aboue 3 times larger, the agreement with the numerical results would
be extended up to $T/\sz$ about 0.8.

We checked that the spatial lattice sizes $\ns$ used in 
figure~\ref{fig:string-tension} 
were large enough for the above conclusions not being altered by any
finite size effects for $T/\sqrt{\sz}\,<\,0.9$ . In practice, below $T/\sqrt{\sz}\approx 0.85$, $m$ is
always larger than about 0.1 and the $\ns$ values used at least 6
times the correlation length $m^{-1}$. In turn, at the largest
$T/\sz$ value shown, the data points exhibit an $\ns$ dependence, to be associated
with effects of the nearby critical temperature (see (\ref{toss})).    
  
At this point, it must be emphasized that this critical temperature 
is astonishingly close to the location of the Nambu-Goto singularity, which is 
from Eq. (\ref{tcng})

\be
\frac{T^{NG}_c}{\sqrt{\sigma_0}}= \sqrt{\frac{3}{\pi}}= 0.977... \label{three}
\ee

\noindent

As a last remark , we point out that while the string model is often supposed to be 
relevant to $SU(N), N\to \infty$ \cite{anthen},
the transition is first order for $N\geq 5$ \cite{svetitsky, liddle},
 and thus cannot be associated with 
the Hagedorn temperature of the string model.

\section{Conclusions} 

\medskip
\noindent
In this article we have described a numerical calculation on the
lattice of the correlations between Polyakov loops in three
dimensional SU(3) gauge theory at finite temperature. We have chosen
$\nt=6$, which we believe is large enough to be in the scaling region
in the three dimensional theory. We vary the temperature by varying
the lattice coupling constant. By doing this we get a dense set of
measurements between $T=T_c/2$ and $T_c$.  We also calculated the
susceptibility of the Polyakov loops in the neighboorhood of the
deconfinement transition, and the moments needed for the Binder method
applied to the variable $|L|^2$.  From this we obtain a determination
of the critical lattice coupling constant. Using results from Teper et
al. we construct an interpolating function and determine the
dimensionless physical quantity $T_c/\sqrt{\sigma_0}=0.976(15)$.  It
is compatible with the Hagedorn temperature of the Nambu-Goto string
model. This is somewhat astonishing, because the string model has the
mean field exponent near the phase transition $\nu=1/2$, whereas from
universality arguments we expect $\nu=5/6$.  To investigate this
further, we use the correlation lengths from the correlation between
Polyakov loops at temperatures below the phase transition, where there
is confinement and the string model may be applicable. The correlation
lengths are inversely proportional to the string tension within the
string interpretation.  We find a very good agreement with the string
tension from the Nambu-Goto string model up to $T\approx 0.7
T_c$. However, up to this temperature the universal terms for a
bosonic string, which coincide with the expansion of the Nambu-Goto
model also describe the data.  Above this temperature we see a
systematic deviation from the string behaviour. This is due to the
fact that here we enter the critical region.

 \medskip
\noindent
In a subsequent paper\cite{bialas2} we will study in detail
 the critical region of the deconfinement transition
in the three dimensional SU(3) gauge theory.

 \medskip
\noindent

\section{Acknowledgments}

\medskip
\noindent
This work is supported by the EU's Sixth Framework Programme, network
contract MRTN-CT-2004-005616(ENRAGE).  B.P. is grateful for the kind
hospitality of the IPhT, CEA-Saclay. B.P. and A.M thank the Institute
of Physics at the Jagiellonian University(Krakow), where part of this
work was performed. One of us (A.M.) is grateful to A. Billoire for
very interesting discussions. The simulations were done on the
computing cluster at Mark Kac Complex System Research Centre.

\end{document}